\documentclass[onecolumn,twoside,aps,superscriptaddress,pre,showpacs,fleqn]{revtex4}
\usepackage{amsmath,amssymb,graphicx}

\newcommand\de{\mathrm{d}}

\newcommand\pa{\partial}
\newcommand\Th{\tilde h}
\newcommand\exh{ h_{\mr{ext}}  }        

\newcommand\TP{ \bigotimes}             

\newcommand\Le{\left}        
\newcommand\Ri{\right} 

\newcommand\rl{ \Ri \rangle}            
\newcommand\lr{ \Le \langle}            

\newcommand\p{ \Le| + \rl}              
\newcommand\m{ \Le| - \rl}              

\newcommand\bp{ \lr  + \Ri|}            
\newcommand\bm{ \lr  - \Ri|}            

\newcommand{\al}{\alpha}                
\newcommand\be{\beta}                   
\newcommand\D{\delta}                   
\newcommand\s{\sigma}                   %
\newcommand\T{\tau}                     %
\newcommand\la{\lambda}                 
\newcommand\La{\Lambda}                 
\newcommand\R{\rho}                     
\newcommand\RR{ {(\rho)}}               

\newcommand\pphi{u}       

\newcommand{\mr}{\mathrm}               
\newcommand{\hs}{\hspace}

\newcommand{\nn}{\nonumber}       

\newcommand{\VB}{ V \hs{-0.07cm} B }    
\newcommand{\SK}{ S \hs{-0.05cm} K }
\newcommand{\LV}{ {Levy} }

\newcommand{\SG}{ S \hs{-0.03cm} G }
\newcommand{\FM}{ F \hs{-0.06cm} M }
\newcommand{\PM}{ P \hs{-0.06cm} M }

\newcommand{\bl}{ \big( }               
\newcommand{\br}{ \big) }

\newcommand\Ph{\mathcal{P}}            
\newcommand\He{\mathcal{H}}            
\newcommand\F{\mathit{f_{trial}}}      
\newcommand\G{\mathbf{g}}              
\newcommand\Pa{\mathbf{\hat \s}}       

\newcommand\sr{;\R \rl \hs{-0.08cm} \rl}
\newcommand\rs{ \Le| \hs{-0.05cm} \Le|}

\newcommand\LS{ \lr \hs{-0.08cm} \lr  }
\newcommand\SL{ ; \R \Ri|\hs{-0.05cm} \Ri|}

\begin{document}
\title{Stability of the replica-symmetric saddle-point in general mean-field
  spin-glass models}
\affiliation{Institut f\"ur Physik, Carl von Ossietzky Universtit\"at,
     26111 Oldenburg, Germany }
\author{Katharina Janzen and Andreas Engel}
\email{janzen@theorie.physik.uni-oldenburg.de , engel@theorie.physik.uni-oldenburg.de}
\date{\today}
\begin{abstract}
Within the replica approach to mean-field spin-glasses the transition
from ergodic high-temperature behaviour to the glassy low-temperature
phase is marked by the instability of the replica-symmetric
saddle-point. For general spin-glass models with non-Gaussian field
distributions the corresponding Hessian is a $2^n\times 2^n$ matrix
with the number $n$ of replicas tending to zero eventually. We
block-diagonalize this Hessian matrix using representation theory of
the permutation group and identify the blocks related to the
spin-glass susceptibility. Performing the limit $n\to 0$ within these
blocks we derive expressions for the de~Almeida-Thouless line of 
general spin-glass models. Specifying these expressions to the cases of
the Sherrington-Kirkpatrick, Viana-Bray, and the L\'evy spin glass
respectively we obtain results in agreement with previous findings
using the cavity approach. 
\end{abstract}
\maketitle
\section{Introduction} \label{intro}
Spin glasses are paradigmatic examples for systems with competing
interactions \cite{BiYo}. Both their equilibrium and dynamical
behavior shows unique characteristics which are absent in  systems
without frustration. The concepts and techniques introduced
in the theoretical description of spin glasses \cite{MPV} have found
interesting and widespread applications in other, at first sight
unrelated fields of science such as complex optimization,
error-correcting codes, artificial neural networks, and computational
complexity \cite{EnvB,HaWe}.  

One of the central features of spin glasses is their non-ergodic low
temperature phase characterized by slow relaxation and
hysteretic response to external magnetic fields. A thorough
theoretical understanding of this phase is available only for
mean-field systems where the spin-glass phase is composed of a hierarchy
of ergodic components. In the parameter plane spanned by temperature
and external magnetic field the high-temperature phase is
separated from the glassy low-temperature phase by the so-called
de~Almeida-Thouless (AT) line \cite{BiYo,AT}. The determination of the
AT-line is therefore of central importance in the theoretical analysis
of spin-glass models.  

Two rather different approaches are by now available to calculate the 
equilibrium properties of mean-field spin glasses. The {\em replica method} 
\cite{EdAn} starts with $n$ replicas of the system under consideration
which after the ensemble average over the quenched disorder
interact with each other. The free energy can be determined from a
saddle-point integral over order parameters. The trademark of the
replica method is the mathematically problematic limit $n\to 0$ to be
performed at the end. In this framework the 
AT-line is determined by the {\it local stability} of the
replica-symmetric saddle point \cite{AT}. In the {\em cavity method}
\cite{ParLH,MPVcav} one spin is added to a system of $N$ spins and the
stochastic stability of the thermodynamic limit $N\to\infty$ is used
to derive self-consistent equations for the order parameters. Here the
AT-line may be obtained by investigating the correlations between 
two spins which must vanish in the thermodynamic limit for a pure
state of a mean-field system \cite{MPV}.  

Both methods have been implemented for the analysis of the simplest 
mean-field spin glass, the Sherrington-Kirkpatrick (SK) model
\cite{SK}. For this model the ergodic phase is characterized
by a single order parameter and a Gaussian distribution of local
magnetic fields. The fluctuations around the replica-symmetric
saddle-point are described by an $n(n-1)/2\times n(n-1)/2$ matrix. Its
eigenvalues have been determined in \cite{AT,PyRu,BrMo}. The 
temperature dependence of these eigenvalues shows that the
replica-symmetric saddle-point loses its stability at the phase
boundary of the ergodic phase. The detailed form of the AT-line was
reproduced within the cavity approach \cite{MPV}.  

The situation is less clear for more general mean-field spin-glass
models which unlike the SK-model are characterized by non-Gaussian
distributions of local fields. Models of this type are in particular
important in complex optimization \cite{RemiNat,HaWe}. A prototype of
this class is the Viana-Bray (VB) \cite{ViBr} model for a diluted spin
glass in which each spin interacts with just a few,  randomly selected
other spins. Here the AT-line was determined numerically in
\cite{JKK}, whereas analytical information is available only near the 
freezing temperature \cite{ViBr}. The replica treatment of diluted spin
glasses and optimization problems is more complicated than that of the
SK-model and involves already at the replica-symmetric level an
infinite number of order parameters \cite{KaSo,MePa}. A general and elegant
approach to this more complicated setting was introduced by Monasson
\cite{Remi}. The fluctuations around the replica-symmetric
saddle-point are now characterized by an $2^n\times 2^n$ matrix which
has to be diagonalized in order to assess the stability of replica
symmetry. Recently it has been shown \cite{JHE} that this method may
also be used to analyze spin-glass models characterized by coupling
distributions with diverging moments such as L\'evy glasses
\cite{CiBo,Cizeau}. This opens up the possibility to determine the AT-line
also for such models within the replica method. 

In the present paper we investigate the stability of the
replica-symmetric saddle-point for spin-glass models with non-Gaussian
field distribution. To this end we implement the approach of Monasson
for diluted spin glasses and reduce the determination of the free
energy per spin to a saddle-point integral over $2^n$ order
parameters. The Hessian matrix describing the fluctuations around this
saddle-point can be block-diagonalized by exploiting the
representation theory of the permutation group \cite{Wigner}. We 
also build on techniques introduced in \cite{WeMo,Martin} for the  
analysis of replica symmetry breaking in one-dimensional spin
glasses. We then identify the blocks which are related to the
spin-glass susceptibility $\chi_{\mathrm{SG}}$ the divergence of which
signals the onset of spin-glass order. Up to this point the analysis
is rather general and uses only the replica structure of the
fluctuation matrix. The final diagonalization of the relevant blocks
can only be performed after the details of the model under consideration
are fixed. We consider three representative examples: the SK model
which merely serves as test case for our method, the VB model as
example for diluted spin glasses, and the L\'evy glass 
as system with a local field distribution exhibiting long tails. In
all cases we provide expressions for the AT-line separating the
replica-symmetric part of the phase space from the region
characterized by replica symmetry breaking.  

The paper is organized as follows. In section \ref{model} we define
the central models of interest, recollect the main steps in the 
replica-symmetric theory for diluted spin glasses and fix the
notation. Section 
\ref{GT} contains the analysis of the situation without external
magnetic field for which the calculations are significantly
simpler. Section \ref{GC} is devoted to the general case from which
the expressions for the complete AT-lines in the models considered
result. Finally, in section \ref{conc} we give some conclusion and
discuss open problems. Some more technical steps are relegated to the
appendices. 

\section{Basic Equations}\label{model}
We consider Ising spins \mbox{$S_i=\pm 1, i=1,...,N$} with random, pairwise
interactions specified by a symmetric matrix $J_{ij}$ in an external
field $\exh$. The Hamiltonian is of the general form 
\begin{eqnarray}\label{ham}
H \Le( \Le\{ S_i \right\}\Ri)=- \frac{1}{2}\sum_{(i,j)}J_{ij}S_iS_j
 - \exh \sum_{i} S_i  \; ,
\end{eqnarray}
where the first sum runs over all pairs of spins. 
The couplings $J_{ij}=J_{ji}$  are i.i.d. random variables drawn from
a model dependent distribution $P_M$. We will consider three examples
for this distribution in detail, namely 
\begin{eqnarray}\label{dist_SK}
P_{\SK}(J)&:=& \sqrt{\frac{N}{4\pi}} \exp \Le( - \frac{ N J^2}{4}\Ri)
\\ \label{dist_VB}
P_{\VB}(J) &:=& \frac{\kappa}{N}\; p(J) -\Le( 1-\frac{\kappa}{N} \Ri) \D(J)
\\ \label{dist_Levy}
\hat P_{\al}(q)&:=&\int  \de J e^{-iJ q} P_{\al}(J)   = \exp\Le(- \frac{|q|^{\al}}{N} \Ri)
\qquad \al \in (0,2]
\end{eqnarray}
corresponding to the Sherrington-Kirkpatrick (SK) 
model \cite{SK}, the Viana-Bray (VB) model \cite{ViBr}, and the L\'evy
spin glass \cite{JHE} respectively. Here $p(J)$ denotes the
distribution of the non-zero bonds in the VB model. The distribution
$P_\al$ for the L\'evy spin glass is defined via its characteristic
function $\hat P_\al$. Note also that the variance of the Gaussian
$P_{\SK}$ is twice its standard value such that it coincides with the
L\'evy case in the  limit $\al\to 2$.

The SK model is the paradigmatic case of a fully
connected spin-glass in which each spin interacts with each other via
weak couplings of order $N^{-1/2}$.
Complementary, the VB model is characteristic for diluted spin glasses 
in which each spin interacts with only a few other spins via strong 
${\cal O}(1)$ couplings.
The L\'evy spin glass interpolates between these two extremes since each
spin interacts with each other spin but most of the couplings are very
weak whereas ${\cal O}(1)$ couplings per spin are strong. 

The large connectivity limit of the VB model leads either  to SK- or  L\'evy-like 
behaviour in the thermodynamic limit, depending on the existence of the second
moment of the distribution  $p(J)$. 
If the second moment exists this limit is defined as
\begin{eqnarray}\label{eq:large_conn_SK}
\kappa \mapsto N, \qquad  p( J) \mapsto \sqrt{N} \,p\Big(\sqrt{N}J\Big)\,,
\end{eqnarray}
and leads to an SK-model  due to the central limit theorem.
On the other hand, if the distribution $p(J)$ has a power law tail:
 $\;p(J) \simeq_{|J| \to \infty}  |J|^{-(\al+1)}, \quad \al \leq 2$,
the large connectivity limit is given by: 
\begin{eqnarray}\label{eq:large_conn_Levy}
\kappa \mapsto N, \qquad p( J) \mapsto  N^{\frac{1}{\al}} p\Big( N^{\frac{1}{\al}} J \Big)\,.
\end{eqnarray}
The limits coincide for $\al = 2 $ as it should be.

In the framework of the replica approach the free energy density $f$,
is expressed through the disorder averaged replicated partition function
$\overline{Z^n(\be)}$ via 
\begin{eqnarray}
f:
=-\lim_{N\to \infty}\frac{1}{\be N}  \overline{ \ln Z \Le( \be \Ri)}
=-\lim_{N\to \infty}\frac{1}{\be N}
   \lim_{n \to 0} \frac{ \overline{ Z^{n}(\be )}-1}{n}\; ,
\end{eqnarray} 
where for integer values of $n$ the replicated partition function is
given by 
\begin{eqnarray}\label{Z^n}
 Z^n\Le( \be \Ri)=\sum_{\{S^a_i \}}
\exp\Le(-\be \sum_{a=1}^n H\Le(\{S^a_i\}\Ri) \Ri) \, .
\end{eqnarray}
After performing the disorder average in (\ref{Z^n}) the trace over
the spin variables can be transformed into a $2^n$-dimensional
integral over order parameters \cite{Remi}
\begin{eqnarray}\label{konz}
c( \vec \s)
=\frac{1}{N}\sum_{i}\delta\Le(\vec S_i, \vec \s \Ri),
\end{eqnarray} 
where $\vec \s = \{ \s_a\}_{a=1\dots n}$ stands for an Ising spin
vector with $n$ components and $\delta(\vec S, \vec \s )$
denotes the Kronecker-$\delta$.
Hence 
\begin{eqnarray}\label{repZ}
\lr Z^{n}(\be) \rl_J
= \int \prod _{\vec \s} \de c \Le( \vec \s \Ri)  
  \delta \Le ( \sum_{\vec \s} c \Le( \vec \s \Ri)-1 \Ri)
  \exp   \Le (-N \be  \F \Le( \Le \{c \Le(\vec \s \Ri)\Ri\} \Ri)\Ri)\; ,
\end{eqnarray}
where the $\delta$-function enforces the constraint 
\begin{equation}\label{constraint}
  \sum_{\vec \s} c \Le(   \vec \s \Ri  )=1
\end{equation}
resulting from (\ref{konz}). The trial free energy $\F$ has two contributions
$\F = f_{S}+f_{E} $ according to 
\begin{eqnarray}  \nn
\be \F (\Le \{ c(\vec \s) \Ri\}) 
&=& 
\sum_{\vec \s}
c\Le( \vec \s \Ri) \ln c \Le(   \vec \s \Ri  )
- 
 \Le[\frac{1}{2}\sum_{\vec \s ,\vec \T} 
c \Le(   \vec \s \Ri  )c \Le(   \vec \T \Ri  )
 \int\de J\,  G_M(J)\,
 \exp \Le( \,\be\,  J\, \vec \s \cdot \vec \T \Ri) 
+ \be \, \exh \,\sum_{\vec \s} c \Le(   \vec \s \Ri  ) \Le(   \sum_{a=1}^n \s_a \Ri  )
\Ri]\; .
\end{eqnarray}
The first term is the entropic contribution $f_{S}$
specifying the number of spin configurations realizing a particular
set of order parameters $c(  \vec  \s )$. The second term $f_E$ derives 
from the Hamiltonian (\ref{ham}) and comprises the interaction energy
and the energy in the external magnetic field. This latter contribution
as well as $f_S$ do not depend on the explicit model considered. 
The interaction part in $f_E$ on the other hand depends on the
specific form of the coupling distribution $P_M$ which is encoded in
$G_M(J)$. For the three models specified in
(\ref{dist_SK})-(\ref{dist_Levy}) one has 
\begin{eqnarray}\label{Gmodels}
G_{\SK}(J)=  {\D}''(J)
\qquad 
G_{\VB}(J)=\kappa \, \big(\, p(J) -\delta(J) \,\big)
\qquad
G_{L}(J)= - \int\, \frac{\de \tilde J}{ 2 \pi} \, | \tilde J|^{\al}\,\exp(\,i\,J \tilde J),
\end{eqnarray}
respectively, where ${\D}'' $ denotes the second derivative of the $\D$-function.
In the  case of the L\'evy spin glass we work at imaginary temperatures $\be =- i k $
as long as  $n \neq 0$ \cite{JHE}.

With the form (\ref{repZ}) the calculation of the free energy
$f$ is reduced to a single site problem as is characteristic
for mean-field systems. In the thermodynamic limit $N\to\infty$ the
remaining integrals over the order parameters can be evaluated
by the saddle-point method. One therefore has to determine the minima
$c_0$ of $\F (\Le \{ c(\vec \s) \Ri\})$ satisfying
\begin{eqnarray}\label{saddle}
c_0\Le(\vec \s \Ri)= \mathcal{L}_{n}
\exp \Le(
\sum_{\vec \T} c_0 \Le(   \vec \T \Ri  )
 \int\de J\,  G_M(J)\,
 \exp \Le( \,\be\,  J\, \vec \s \cdot \vec \T \Ri) 
+ \be \, \exh  \sum_{a=1}^n \s_a 
         \Ri) \; ,
\end{eqnarray}
where $\mathcal{L}_{n}$ is a factor accounting for the constraint
(\ref{constraint}). 

The general solution of the saddle-point equation is a complicated
problem. To get some guidance one uses as a first step the assumption 
of replica symmetry (RS) stipulating that $c_0$ depends on $\vec \s$
only through the sum $\sum_{a=1}^n\s_a$ of the vector
components. These replica-symmetric order parameters are related
to the (replica-symmetric) distribution of local fields $\Ph (h)$ in the
spin glass by \cite{Remi} 
\begin{eqnarray}\label{RS}
c_0\Le(\vec \s \Ri)
=  c\Le(  \sum_{a=1}^n\s_a \Ri )
= \int \de h  \, \Ph (h) \, \frac{\exp(\be h \sum_{a=1}^n\s_a ) }{ \Le (2 \cosh (\be h)
  \Ri)^n }\; . 
\end{eqnarray}

As is well known the replica-symmetric solution fails at low
temperatures which on the formal level is due to the instability of 
the replica-symmetric saddle-point. To assess this stability the
temperature dependent eigenvalues of the Hessian $\He$ describing the
quadratic fluctuations around the replica-symmetric saddle-point have
to be determined. The matrix elements of $\He$ are given by 
\begin{eqnarray}\label{He}
\He \Le(\vec \s, \vec \T \Ri)&=&
\Le. \frac{ \be \,\,  \pa^2 \F}{ \pa c\Le( \vec \s \Ri) \pa c \Le( \vec \T \Ri) } \Ri|_{RS}
=\frac{\D_{\vec \s, \vec \T }}{ c\big(\sum_{a=1}^n\s_a \big)}- 
 \int \de J \, G_M(J)\, \exp \Le( \,\be\,  J\, \vec \s \cdot \vec \T \Ri).
\end{eqnarray}
Similarly to $\F$ the Hessian splits into an entropic and an energetic
contribution, $\He = \He_S + \He_E $. Note that the external field
does not show up explicitly in $\He$. Nevertheless it influences the
stability of the replica-symmetric saddle-point due to the dependence
of $c_0$ on $\exh$ as specified by (\ref{saddle}). Note
also that the expression (\ref{He}) does not yet account for the
constraint (\ref{constraint}). 

The determination of the eigenvalues of
$\He$ may be systematically simplified by exploiting the permutation
symmetry of the replica-symmetric saddle-point. To this end it is
convenient to express the Hessian as a tensor product of $2\times 2$
matrices using  
\begin{eqnarray}\label{tensor_decomp_energ}
e^{\be J\, \vec \s \cdot \vec \T} 
=\prod_{a=1}^n e^{\be  J\,  \s_a  \T_a}
=\prod_{a=1}^n 
\lr \s_a \Ri|
\Le( e^{\be  J} \Pa_0+e^{-\be  J} \Pa_1 \Ri)
 \Le| \T_a \rl
=\lr \vec \s \Ri|
 \TP_{a=1}^n 
 \Le( e^{\be  J} \Pa_0+e^{-\be  J} \Pa_1 \Ri)
 \Le | \vec \T\rl 
\end{eqnarray}
for the energetic and 
\begin{eqnarray}\label{tensor_decomp_entrop}
\frac{\D_{\vec \s, \vec \T }}{  c(\sum_{a=1}^n\s_a     )}
=\int \frac{\de r \de r' }{2 \pi}  \frac{e^{irr'} }{  c(r)}
\prod_{a=1}^{n}  \D_{  \s_a ,\T_a} e^{-ir'\s_a} 
=\lr \vec \s \Ri|
 \int \frac{\de r \de r' }{2 \pi} \frac{e^{irr'}}{  c(r)}
 \TP_{a=1}^n \exp \Le(-ir' \Pa_3  \Ri)
 \Le | \vec \T\rl 
\end{eqnarray}
for the entropic part, respectively. Here the vectors $\Le | \vec \s
\rl$  are  defined as $  \Le | \vec \s   \rl=\Le| \s_1, \s_2 , \dots
\s_n \rl  = \bigotimes_{a=1}^{n}  \Le | \s_a   \rl$ with $\s_a=\pm1$
which span the space $V$ of replicated spin configurations. The
matrices 
\begin{eqnarray}
\Pa_0=\Le(\begin{array}{cc}1&0\\0&1\end{array}\Ri) \, ,  \quad
\Pa_1=\Le(\begin{array}{cc}0&1\\1& 0\end{array}\Ri)\, , \quad
\Pa_3=\Le(\begin{array}{cc}1& \hs{+0.1 cm }  0\\0& \hs{-0.1 cm }-1\end{array}\Ri)  \, .
\end{eqnarray}
are the usual Pauli matrices.
\section{Zero external field}\label{GT}
The discussion of the eigenvalues of the Hessian $\He$ is relatively
straightforward for $\exh=0$ and $T \geq T_c$. We therefore first derive
expressions for the eigenvalues in this region and afterwards turn to
the more involved situation with $\exh\neq 0$. 

For $\exh=0$ the saddle-point equation (\ref{saddle}) has for all
$\beta$ the solution 
\begin{equation}\label{cpara}
  c \Le ( \sum_{a=1}^n\s_a  \Ri)=\frac{1}{2^{n} }
\end{equation}
which, according to  (\ref{RS}), corresponds to $P(h)=\delta(h)$. On
physical grounds we expect that this paramagnetic solution is stable at
sufficiently high temperatures. For constant $c(s)$ the 
Hessian (\ref{He}) for the paramagnetic solution reads
\begin{eqnarray}
 \lr \vec \s \Ri| \He_{\PM }  \Le | \vec \T\rl= 
  \lr \vec \s \Ri| \,2^n \TP_{a=1}^n \Pa_0 -\int \de J  \, G(J) 
 \TP_{a=1}^n
\Le( e^{\be  J} \Pa_0 +e^{-\be  J} \Pa_1  \Ri)   \Le | \vec \T\rl\; .
\end{eqnarray}
Its eigenvectors can therefore be constructed from those of
$\Pa_1$, namely
\begin{eqnarray}
| k \rangle = \frac{1}{\sqrt{2}}  \Big(  \Le|+ \rl +( -1)^k \Le |- \rl \Big) \qquad k \in \{0,1\}\,.
\end{eqnarray}
The eigenvectors of $\He_{\PM}$ may hence be written in the form 
\begin{eqnarray}
  | \vec k \rangle &= & \Le| k_1,\dots ,k_n \rl  =\TP_{a=1}^{n}  \Le|
  k_a \rl  
  \qquad \text{with}\qquad \Le| k_a \rl   \in 
  \Le \{\phantom{A^k_µ} \hs{-0.5cm}   
  \Le|1 \rl ,   \Le|0 \rl 
  \Ri \}\; .
\end{eqnarray}
The $2^n$ mutually orthogonal vectors $| \vec k \rangle$
form a basis in the space $V$. With the abbreviation $k=\sum_a k_a$ 
we find  
\begin{equation}
   \TP_{a=1}^n \Le( e^{\be  J} \Pa_0 +e^{-\be  J} \Pa_1 \Ri)
   |\vec k\rangle =(2\cosh(\be J))^{n-k}(2\sinh(\be J))^k \;|\vec k\rangle 
\end{equation}
and therefore the eigenvalue corresponding to $| \vec k \rangle$ is
given by 
\begin{eqnarray}
\La^{\vec k }_{n}
&=&2^n \Le( 1-\int \de J \, G_M(J) \,\cosh^n(\be  J) \tanh^{k}\Le(\be  J
\Ri)  \Ri)  \; .
\end{eqnarray} 
Not all of these eigenvalues are, however, relevant for the stability
of the saddle-point since the constraint (\ref{constraint}) needs
still to be taken into account. To do so consider small fluctuations
$\delta c(\vec \s)$ around the RS saddle-point $c_0(\vec
\s)=2^{-n}$. With the notations 
$\delta c(\vec \s)=\langle \delta c |\vec \s\rangle$ and 
$\delta \hat c(\vec k)=\langle \delta c | \vec k\rangle$ we  find
from (\ref{repZ}) to second order in $\delta c(\vec \s)$ 
\begin{eqnarray}\label{irrDir}  
\overline{ Z^{n}(\be)}
&\approx&
\exp\big(-N\be  \F(\{2^{-n}\}) \big) \int  \prod _{\vec \s} \de \D c \Le( \vec \s \Ri)
 \,\delta \Le( \sum_{\vec \s} \D c \Le( \vec \s \Ri) \Ri)
\exp \Le(-\frac{N}{2}  \sum_{\vec  \s  ,  \vec \T  } \D c (\vec  \s)
\He_{\PM} (\vec \s,\vec \T)    \D c (\vec \T) \Ri)\\ \nn 
&=&\exp(-N \be \F(\{2^{-n}\})) 
\int  \prod _{\vec k} \de \D  \hat c ( \vec k ) 
\,\D \Le( 2^{\frac{n}{2}} \D  \hat c(\vec 0 ) \Ri)
\,\exp\Le( -\frac{N}{2}\,\sum_{\vec k} \,\La_{ \vec k} \, \D \hat
c(\vec k )^2  \Ri) \\ \nn 
&=& \exp\Le( -N \be \F(\{2^{-n}\}) -\frac{n}{2}\ln 2 \Ri ) 
\int  \prod _{\vec k \neq \vec 0}  \de \D  \hat c ( \vec k ) 
\,\exp \Le( -\frac{N}{2}  \, \sum_{\vec k \neq  \vec 0}\, \La_{ \vec
  k}\,  \D \hat  c(\vec k )^2  \Ri) \; .
\end{eqnarray} 
Here we have used the transformations 
\begin{eqnarray}\nonumber
\sum_{\vec \s} \D c \Le( \vec \s \Ri)
= \sum_{\vec \s} \lr \D c \Ri| \Le. \hs{-0.05 cm} \vec \s \rl 
= \sum_{\vec k} \lr \D c \Ri|  \vec k \rangle  \sum_{\vec \s}  \langle
\vec  k \Le| \vec \s \rl  
=\sum_{\vec k}  \D \hat c(\vec k) \,   2^{\frac{n}{2}}  \, \D_{\vec k , \vec 0}   
=2^{\frac{n}{2}}  \, \D \hat c(\vec 0)  
\end{eqnarray} 
and 
\begin{eqnarray}\nonumber
\sum_{\vec \s ,\vec \T}  \D c \Le( \vec \s \Ri) \He(\vec \s, \vec \T )
\D c \Le( \vec \T \Ri) 
=  \sum_{\vec k} \, \La^{\vec k }_{n} \, {\D \hat c(\vec k) }^2.
\end{eqnarray} 
Consequently  $ |\vec 0 \rangle$ is perpendicular to the constraint
(\ref{constraint}) and the  integration in the corresponding direction 
is suppressed by the $\D$-function in (\ref{repZ}). The value of
$\La^{\vec 0}_n$ is therefore not relevant for the convergence of the
integral (\ref{irrDir}).

The limit $n \to 0$ can be performed now for every eigenvalue
$\La^{(k)}_{n}:=\La^{\vec k}_n  $ corresponding to the eigenspace spanned by
eigenvectors with magnetization $k$
\begin{eqnarray}\label{PMEW}
\La^{(k)}_{\PM} = \lim_{n \to 0} \La^{\vec k}_n =
1-\int \de J \, G_M(J) \tanh^{k}
\Le(\be  J \Ri) \,  .
\end{eqnarray}
%
%
The SG transition is signaled by the divergence of the SG susceptibility
given by \cite{BiYo}
\begin{eqnarray}\label{chi}
\chi_{\SG}&:=&\frac{1}{N}\sum_{i,j} 
 \overline{\lr S_i S_j \rl ^2_c }
=\frac{1}{N} \sum_{i,j} \lim_{n \to 0}
\frac{1}{n(n-1)}\sum_{(a,b)}\lr S^{a}_i  S^{a}_j S^{b}_i
S^{b}_j   \rl_{\rm{repl}}. 
\end{eqnarray}
where the second equality holds only in the paramagnetic phase.
The last average is to be taken with respect to all distinct replicas
after the disorder average has been performed and yields 
\begin{eqnarray}\label{repl_av}
\lr S^{a}_i  S^{a}_j S^{b}_i  S^{b}_j   \rl_{\rm{repl}}
= \sum_{ \Le \{ S^{a}_i \Ri \}}  
S^{a}_i  S^{a}_j S^{b}_i  S^{b}_j 
\exp
\Le(
\frac{1}{2N} \sum_{i,j}\int \de J \, G_M(J) \,  e^{ \be J  \vec S_i \cdot \vec S_j}+
 \be \exh  \sum_{i=1}^N\sum_{a =1 }^n S_i^a
\Ri) \,.
\end{eqnarray} 
Using the same method as for the replicated partition function $\overline{ Z^n(\be )} $ 
it is  possible to rewrite (\ref{repl_av}) as an $2^n$-dimensional
integral. 
In the  saddle-point approximation one finds using $ c_{0}(\vec \s ) = 2^{-n}$
\begin{eqnarray}\nn
  \chi_{\SG}&=&N+ \lim_{n \to 0} \frac{N}{n(n-1)} \int  \prod _{\vec
    \s} \de c \Le( \vec \s \Ri)  
  \delta \Le ( \sum_{\vec \s} c\Le( \vec \s \Ri)
-1 \Ri)
  \sum_{\vec \s ,\vec \T} c\Le(\vec \s \Ri) 
  c\Le (\vec \T \Ri) 
  \Le(\vec \s \cdot \vec \T \Ri)^2
  \exp \big(-N \be  \F\Le( \Le \{c\Le(\vec \s \Ri)
\Ri\} \Ri)\big)\; \\ \nn
&& \hs{-1cm}\approx
 \lim_{n \to 0} \frac{N
e^{-N\be  \F(\{2^{-n}\})}   }{n(n-1)} \!\!
\int  \prod _{\vec \s} \de \D c \Le( \vec \s \Ri)
\delta \Le( \sum_{\vec \s} \D c \Le( \vec \s \Ri) \Ri)
 \sum_{ \vec  \s  ,  \vec \T  }  \D   c \Le (\vec  \s \Ri) 
 (\vec  \s  \cdot  \vec
\T   )^2  \D c(\vec \T)  
\exp \Le( \!\! -\frac{N}{2}   \sum_{\vec  \s  ,  \vec \T  }
 \D   c (\vec  \s) 
\He_{\PM} (\vec \s,\vec \T)    \D c (\vec \T) 
\Ri) \\ \nn
&=&\frac{1}{\La^{(2)}} .
\end{eqnarray} 

The spin-glass susceptibility is hence directly related to one of the
eigenvalues of the Hessian. As expected the divergence of $\chi_{SG}$
at the spin-glass transition corresponds to the instability of the
replica-symmetric saddle-point signaled by $ {\La^{(2)}} =0$. 
%
%
We also mention that for non-symmetric coupling distributions a
transition to a ferromagnetic phase may occur. This transition is
related to the ferromagnetic susceptibility 
\begin{eqnarray}
\chi_{\FM}&
:= &\frac{1}{N}\sum_{i,j} 
\overline{  \lr S_i S_j \rl_c    }
=\frac{1}{N} \sum_{i,j} \lim_{n \to 0} \frac{1}{n}\sum_{a}\lr S^{a}_i  S^{a}_j \rl_{repl} 
= \frac{1}{ \La^{(1)}}
\end{eqnarray} 
and its divergence is hence linked with another eigenvalue of $\He_{\PM}$. 
\subsection*{Examples}
We now discuss the results for the three particular models defined
before. For the SK-model we find from (\ref{PMEW}) 
\begin{eqnarray}
\La_{\SK}^{(k)}= 1- 2 \,\D_{2,k}\,\be^2\, .
\end{eqnarray}
In accordance with the symmetry $P_{SK}(-J)=P_{SK}(J)$
(cf.~(\ref{dist_SK})) only a transition to a spin-glass phase is
possible at low temperature. Defining the transition temperature $T_c$
by $\La_{\SK}^{(2)}=0 $ we find $T_c=\sqrt{2}$ and 
\begin{eqnarray}\nn
  \chi_{\SG}=\frac{T^2}{T^2-T_c^2}
\end{eqnarray} 
which coincides with the well-known results \cite{BiYo}.

The expression for the eigenvalues of the Hessian for the VB-model  
\begin{eqnarray}\label{VBEW}
\La^{(k)}_{\VB} = 1- \kappa \int \de J \, p(J) \, \tanh^{k}
\Le(\be  J \Ri)+ \kappa \, \D_{k,0}  
\end{eqnarray}
was already derived by Monasson \cite{Remi_priv_comm}. Since
$\tanh(\be J)\leq 1$ the integrals become smaller with increasing
$k$ while the fluctuations corresponding to $k=0$ are suppressed
by the constraint (\ref{constraint}). For symmetric distributions
$p(J)$ the integral for $k=1$ vanishes identically and hence
$\La^{(2)}_{\VB}$ is the first eigenvalue to become negative. From
$\La^{(2)}_{\VB}=0$ results the well-known expression \cite{ViBr} for
the critical temperature of the spin-glass transition in the
VB-model \cite{ViBr,KaSo,MePa}: 
\begin{eqnarray}
1=\kappa \int \de J\, p(J)\, \tanh^2(\be_c J)
\end{eqnarray}

For the L\'evy glass we have 
\begin{eqnarray}\label{LEW}
\La^{(k)}_{\LV} = 1+\int \frac{\de J \, \de \tilde J}{ 2 \pi} 
   \, | \tilde  J|^{\al}\,\exp(\,i\,J \tilde J)\, \tanh^{k}\Le(\be  J \Ri) \,.
\end{eqnarray}
Again $P_{\al}(J)=P_{\al}(-J)$ ensures that 
\begin{equation}\label{eq:revJJ'}
  \La^{(2)}_{\LV}= 1+\int \frac{\de J \de \tilde J}{ 2 \pi} 
   \, | \tilde  J|^{\al}\,\exp(\,i\,J \tilde J)\, \tanh^2\Le(\be  J \Ri)
   =1-\frac{\Gamma(\alpha+1)}{\pi} \sin \Le( \frac{\alpha\pi}{2} \Ri)
       \int \frac{\de J}{|J|^{\al+1}} \tanh^2\Le(\be  J \Ri)
\end{equation}
is the first eigenvalue to become negative. For the critical
temperature of this spin-glass transition we hence find 
\begin{eqnarray}\label{defC}
T^\al _c= C(\al)
  \int \frac{\de J}{|J|^{\al+1}}\tanh^2 (J) \qquad \mr{with}\quad  C(\al)= \frac{\Gamma(\alpha+1)}{\pi}\sin \Le ( \frac{\alpha\pi}{2} \Ri)
\end{eqnarray}
which was previously derived by both the cavity \cite{CiBo} and the
replica method \cite{JHE}. For the SG-susceptibility of this model
results  
\begin{equation}\nn
  \chi_{\SG}=\frac{T^\al}{T^\al-T_c^\al}
\end{equation} 
which is similar to the expression in the SK-model and in fact  
comprises it for $\alpha=2$.  

We therefore reproduce for all three examples the known results about
the instability of the paramagnetic phase. However, it 
remains unclear at this point whether these instabilities can be cured
within the replica-symmetric sector or whether replica symmetry
breaking is necessary to stabilize the saddle-point. This question can
only be addressed by investigating the more complicated situation with
a non-trivial form of $c \big(\sum_a \sigma_a \big)$ to which we turn
now. 

\section{The general case}\label{GC}
At low temperatures or $\exh\neq 0$ the Hessian $\He$ explicitly
depends on $\Pa_3$ and the determination of its eigenvalues becomes
much more complicated. The essential steps of the analysis are as
follows: We first use the permutation symmetry between replica indices
characteristic for any RS solution to block-diagonalize the Hessian
according to the irreducible representations of the permutation
group. For the remaining diagonalization inside the blocks the limit
$n\to 0$ can be performed and the eigenvalue equations assume the form
of integral equations. We then show that the replicon eigenvalue
related to the spin-glass susceptibility can only belong to one of the
first three blocks. By an expansion around $T_c$ we then verify that
all eigenvalues of the zeroth and first block which go to zero at
$T_c$ return to positive values below $T_c$. Consequently the replicon 
eigenvalue must lie in the second block. Finally we identify this
eigenvalue and give explicit expressions for the AT-line for the three
special models considered. 

A major simplification of the general eigenvalue problem of the
Hessian is obtained by invoking the symmetry of $\He$ under
permutations of the replica indices. Formally this symmetry is
expressed by the commutation of the Hessian with a representation $D$
of the permutation group $\Sigma_n$.  In our case $D$ is defined by    
\begin{eqnarray} \nn
  D(\pi)\left |\s_{1},\s_{2}\dots \s_{n} \rl =\left
      |\s_{\pi(1)},\s_{\pi(2)}\dots \s_{\pi(n)} \rl  
\qquad  \left |\s_a\rl  \in \big \{ \left | +\rl,\left |-\rl \big \} \qquad \pi \in \Sigma_n   ,
\end{eqnarray}
which clearly commutes with $\He$ , i.e.
\begin{eqnarray}\label{commut}
D(\pi)\He = \He D (\pi) \qquad \forall \pi \in \Sigma_n .
\end{eqnarray}
The Hessian can  therefore be block-diagonalized according to the irreducible
representations $D^{(\R)}$ of $\Sigma_n$ that are contained in $D$.
As a first step of the reduction of $D$ we note that the subspaces
$V^{(\s)} $, spanned by vectors $\Le |\s_{1},\s_{2}\dots \s_{n} \rl  $  
with a fixed number $\s $ of entries  $\s _a$ equal to plus one,  are
invariant under $D$. The restricted  representation $\Delta^{(\s)}$ of
$D$ on the subspace $V^{(\s)}\;\,  \s =0 \dots  n $ can be further
decomposed into irreducible representations: 
\begin{eqnarray}\label{D-decomp}
\Delta^ {(\s)} \simeq D^{(0)}\oplus  D^{(1)} \dots  \oplus  D^{(\tilde \s)} 
\end{eqnarray}
with $ \tilde \s= \mr{min}\{ \s, n-\s \} $. The irreducibility of the 
$D^\RR$'s is shown in  \cite{Wigner}.  
Each representation $D^{(\R)}$ gives rise to an eigenvalue
$\la^{(\R)}$ of $\He$, with degeneracy  
\begin{eqnarray} 
\mr{deg}\Le(\la^{(\R)} \Ri)= \mr{dim} \Le( D^{(\R)} \Ri )
 ={n \choose \R}-{n \choose \R-1}.
\end{eqnarray}
The  subspaces $V^{(\s)}$ hence  split into direct sums of
irreducible subspaces $V^{(\s,\R)}$ each of which is associated with a
representation $D^{\RR}$, 
\begin{eqnarray}
V^ {(\s)} \simeq V^{(\s,\, 0)}\oplus  V^{(\s,\, 1)} \dots  \oplus
V^{(\s, \,\tilde \s)} \,. 
\end{eqnarray}
This decomposition can be accomplished by  Young-symmetrizers
constructed using the Young tableaus \cite{Weyl}. 
The procedure shows that the vectors 
\begin{eqnarray} \label {unsym}
\rs  \s  \sr
:=   {\big(\p \m \big)}^{\R}
    {\p}^{\s-\R}{\m}^{n-\s-\R}\,  \in  V^{(\s)}, 
\quad  \R = 0 \dots \tilde \s  \,,
\end{eqnarray}
are mapped onto the irreducible invariant subspace $V^{(\s,\R)}$ by
an anti-symmetrization in two successive entries in the first $2\R$ arguments, 
and a complete symmetrization in the last $n-2\R$ entries.
The vector
\begin{eqnarray}\label {sym}
\Le |  \s ; \R  \rl  &:=& \mathcal{A}_{2\R} \,\mathcal{S}_{n-2\R}(\s)
\rs  \s  \sr
 = \big( \p \m-\m\p \big)^{\R}\Le|   \s -\R \rl_{n-2 \R}   
\end{eqnarray}
hence lies in $ V^{(\s, \R)}$, where the operators $\mathcal{A}_{2\R}$
and $\mathcal{S}_{n-2\R}(\s)$ denote the anti-symmetrization and the
symmetrization operators, respectively and the symmetrized part of the
vector $\Le |  \s;\R  \rl$ with $\sigma-\R$ entries equal to plus one
is denoted by 
\begin{eqnarray} \label{sympart}
\Le|  \s -\R \rl_{n-2 \R}= \sum_{\sum \s_a =  2 \s-n}^{}
\TP_{a=1}^{n-2 \R} \Le|\s_a \rl.   
\end{eqnarray}

A basis of the subspace $ V^{(\s, \R)}$ can be constructed by applying 
all the $D(\pi)$ on $\Le |\s;\R\rl$ and choosing a maximal linearly
independent subset. We note that the vectors  $\{ \Le |  \s ;\R \rl,
\hs{0.2cm} \s=0\dots n, \hs{0.2cm}\R=0 \dots \tilde \s \}$ are
orthogonal, but not normalized. For a fixed $\R$  the set of the
normalized vectors  
\begin{eqnarray}
\Le\{  \frac{\Le |  \s ;\R \rl}{ \sqrt{ \lr \s;\R   \Ri. \Le|  \s  ;\R\rl } }, \hs{0.5cm}  \s= \R\dots n- \R  \,  \Ri  \}
\end{eqnarray} 
is an orthonormal basis of a subspace $W^\RR$ of $V$, containing one
element from each irreducible subspace 
\mbox{$ V^{(\s,\R)},  \hs{0.2cm}\s=\R\dots n-\R$}. 
The matrix $\He^\RR_{\rm Sym}$ with  matrix elements 
\begin{eqnarray}\label{rhoBlock}
\He^\RR_{ \rm Sym}(\s ,\T )
& =& 
\frac{  \lr   \s ; \R \Ri | \He \Le|   \T;\R  \rl }
{ \sqrt{   \lr   \s; \R  \Ri | \Le.   \s ; \R \rl   \lr   \T; \R  \Ri
    | \Le.   \T  ;\R \rl   }  } 
\qquad \s , \T = \R \dots  n-\R 
\end{eqnarray}
therefore comprises information from each irreducible subspace 
\mbox{ $V^{(\s,\R)},\hs{+0.2cm} \s=\R\dots n-\R  $} 
\cite{Pendry, WeMo,Martin} .
Diagonalization of the $n+1-2\R$ dimensional matrix $\He^\RR_{ \rm
  Sym}$ leads to the eigenvalues
\begin{eqnarray}
{\la^\RR_\s,\qquad \s= \R \dots n-\R }
\end{eqnarray}
each of which is associated with one representation  $D^{\RR}$  arising in the
decomposition of $D$.

Exploiting the symmetry of the problem we have hence reduced the $2^n$
dimensional eigenvalue problem to  $\Le \lfloor \frac{n}{2} \Ri \rfloor$ 
eigenvalue equations of dimension \mbox{$n+1-2\R$} which are
parametrized by $\R $. Here $\Le \lfloor x \Ri \rfloor$ denotes 
the largest integer smaller than $x$. 
As shown in appendix \ref{App:n_to_0} the permutation symmetry of $\He$ 
can be further used to switch from the eigenvalue problem with symmetric 
matrices $\He^\RR_{\rm Sym}$ to non-symmetric matrices with matrix elements 
$\He^\RR(\s,\T)=\LS \s \SL \He\Le| \T ; \R  \rl$. 
This form allows an elegant continuation $n \to 0$ which was
already used in \cite{WeMo,Martin}. 
We expect that the eigenvalues of the matrices  $\He^{\RR}$ are
generically non-degenerated since the permutation symmetry is 
already completely accounted for.

Finally the eigenvalues $\la^\RR$ corresponding to the representations
$D^{\RR}$ have to be determined from the eigenvalue equation:
\begin{eqnarray}\label{EV1}
\la^{(\R)}   \chi^{(\R)}(\T) 
= \sum_{\s =\R}^{n-\R} \He^{(\R)}(\s ,\T) \chi^{(\R)}(\s) 
\end{eqnarray}
The decomposition $\He = \He_S + \He_E$ of the Hessian still 
holds in the new basis. The entropic part $\He_S$ is diagonal and
depends on the RS solution  $c\bl\sum_a \sigma_a\br$ of the saddle-point
equation only: 
\begin{eqnarray}\label{HeRhoS}
\He^\RR_{S}(\s, \T) = \frac{\delta_{\s, \T}}{ c(2\s-n)}\; .
\end{eqnarray}
The energetic part $ \He^\RR_E$ depends on the details of the
Hamiltonian and will be specified in the treatment of the particular
models below. For the following analysis it is convenient to multiply
the eigenvalue equations  with the inverse of $\He^\RR_S$ and to
transform the eigenvectors 
$\chi^\RR $ to functions $\phi^\RR$ according to \cite{WeMo,Martin} 
\begin{eqnarray}\label{n20}
\phi^\RR(h):= \sum_{\s=\R}^{n-\R}\chi^\RR(\s) \exp \bl \be h (2\s -n) \br\,.
\end{eqnarray}
The limit \mbox{$n\to 0$} may then be performed which transforms the
finite dimensional matrix equations into integral equations. In the
case of the L\'evy glass it also allows the  continuation to real
temperature. Some intermediate steps of the calculations are given in
appendix \ref{App:ev_eq}. 

Eventually we arrive at eigenvalue equations of the form
\begin{eqnarray}\label{EV2}
\la^\RR \,
\int \de \Th \, \Ph(\Th)\,\phi^\RR \big(\Th+h \big)
&=&\phi^\RR( h ) 
-\int \de \Th \, \Ph (\Th) 
 \int \de J \, G_M(J) \,
\Le (\pa_h \pphi \Ri )^\R( \Th+h,  J)\,
\phi^\RR \big( \pphi( \Th+h,  J) \big)
\end{eqnarray}
where $ \pphi(h,J)= \frac{1}{\be}\mr{atanh}\big( \tanh (\be h ) \tanh( \be
  J)\big)$ and $\Ph (h)$ denotes as before the replica-symmetric 
distribution of local fields. This equation is the central
  result of the present section.

As a first test we reproduce the spectrum of
the Hessian obtained in section \ref{GT} for $\exh=0$ and $T \geq T_c$. In
this case  $\Ph(h)=\delta(h)$ and the eigenvalue equations simplify to  
\begin{eqnarray}\label{PMEW2}
\la^\RR \,
\phi^\RR(h )
&=&\phi^\RR( h ) 
- \int \de J \, G_M(J) \,
\Le (\pa_h \pphi \Ri )^\R( h,  J)\,
\phi^\RR( \pphi( h,  J) ). 
\end{eqnarray}
Setting $h=0$ and using $\pphi(h=0,J)=0$ as well as $\pa_h \pphi
(h=0,J)=\tanh \Le( \beta J \Ri)$ we get 
\begin{eqnarray}\label{PMEW3}
\la^\RR \,
\phi^\RR(0)
&=&\phi^\RR( 0 ) 
- \int \de J \, G_M(J) \,
\tanh^\R(\beta J)\,
\phi^\RR(0). 
\end{eqnarray}
If $\phi^\RR(0)\neq 0$ we hence find $\La^\RR=\la^\RR$ and therefore
reproduce expression (\ref{PMEW}) for the eigenvalues obtained more
directly in section \ref{GT}. If on the other hand $\phi^\RR(0)=0$,
then eq.~(\ref{PMEW3}) 
does not convey any information about $\la^\RR$. However, in this case we
find after differentiating (\ref{PMEW2}) with respect to $h$ and
setting $h=0$ afterwards  
\begin{eqnarray}\label{PMEW4}
\la^\RR \, \bl \phi^\RR\br'(0)&=&\bl \phi^\RR\br'( 0 ) 
- \int \de J \, G_M(J) \,\tanh^{(\R+1)}(\beta J)\,\bl \phi^\RR\br'(0) \,. 
\end{eqnarray}
If $ \Le( \phi^\RR \Ri)'(0)\neq 0$ this returns the expression for
$\La^{(\R+1)}$. If $\Le( \phi^\RR\Ri)'(0)= 0$ we turn to the second
derivative of (\ref{PMEW3}) and so on. 
In conclusion we find that for $\exh=0$ and $T \geq T_c$ the
sub-block $\He^{\RR}$ defined in (\ref{rhoBlock}) generates all
eigenvalues $\La^{(k)}$ from (\ref{PMEW}) with $k\geq\rho$.
Conversely the eigenvalue $\La^{(2)}$ which signals the spin-glass transition 
shows up only in blocks $\R = 0 , 1, 2$ implying that in some
neighbourhood of $T_c$ all eigenvalues $\la^{\RR}$ with $\R>2$ are
strictly positive. This is also corroborated by a replica
representation of the spin-glass susceptibility starting with
(\ref{chi}) which shows that the SG susceptibility does not depend on
eigenvalues $\la^{\RR}$ with $\R > 2$. In the following we therefore
investigate only the ``dangerous'' blocks $\R=0,1,2$.

For $\R=0$ the constant function  $\phi^{(0)}(h)=1$ is an
eigenfunction corresponding to the eigenvalue $\lambda^{(0)}=1= 
\Lambda^{(0)}$. In the high temperature region this eigenvalue was
irrelevant for the stability due to the constraint (\ref{constraint}).
We assume that the same holds true in the spin-glass phase as well. 
For the first derivative of (\ref{EV2}) we find for \mbox{$(\R=0)$} 
\begin{eqnarray}
\la^{(0)} \,
\int \de \Th \, \Ph(\Th)\, \bl \phi^{(0)}\br'(\Th+h )
&=& \bl \phi^{(0)}\br'( h ) 
-\int \de \Th \, \Ph(\Th) 
 \int \de J \, G_M(J) \,
\pa_h \pphi ( \Th+h,  J)\,
\bl \phi^{(0)} \br '\bl \pphi( \Th+h,  J) \br.
\end{eqnarray}
Hence either \mbox{$ \bl \phi^{(0)} \br'\equiv 0 $} or $\bl
\phi^{(0)}\br'$ is an eigenfunction of (\ref{EV2}) for 
\mbox{$\R = 1$}. Conversely if $\phi^{(1)}$ is an eigenfunction of
(\ref{EV2}) for $\R=1$ its primitive $\Phi^{(1)}$ satisfying $ \bl
\Phi^{(1)} \br'= \phi^{(1)} $ fulfills the equation
\begin{eqnarray}
\partial_h \left\{ \la^{(1)} \, \int \de \Th \, \Ph(\Th)\,  \Phi^{(1)}(\Th+h )
 \right\} &=& \pa_h \left\{ \Phi^{(1)}( h ) -\int \de \Th \, \Ph(\Th)  
\int \de J \, G_M(J) \, \Phi^{(1)}\bl \pphi( \Th+h,  J) \br \right\}\,.
\end{eqnarray}
Integration of this equation yields an eigenfunction $\Phi^{(1)}$ of
(\ref{EV2}) for $\R =  0$ since the integration constant may be
absorbed in the choice of $\Phi^{(1)}$. Consequently the
block with $\R=0$ contains the same eigenvalues as the block
corresponding to $\R=1$ and in addition one eigenvalue corresponding
to a constant eigenfunction which we believe to be irrelevant due to
the constraint (\ref{constraint}). This degeneracy between the $\R=0$
and the $\R=1$ block is similar to the well-known degeneracy of the
longitudinal and first transversal eigenvalue in the stability
analysis of the SK model \cite{AT}. 

We now show that the eigenvalues $ \la^{(0)}$ and $ \la^{(1)}$ which 
are degenerate with  $\La^{(2)}$ in the high temperature phase return
to positive values below $T_c$. In view of the equivalence between the
eigenvalues from the zeroth and first block it is sufficient to show
this for $ \la^{(1)}$.
We study the eigenvalue equation perturbatively to leading order 
in the reduced temperature $\tau= 1-T/T_c$ at zero external field.
To this end we expand the derivative of the $\R=1$ eigenvalue equation with
respect to $h$ at $h=0$ up to order $h^2$ \cite{JEM1}. The integral equation acquires
the form of a $2$-dimensional matrix eigenvalue problem. To leading order
in $\tau$ we find 
\begin{eqnarray} \label{eq:ev_rho_1}
\la^{(1)}_{\SK}=2 \,        \tau\; , \qquad 
\la^{(1)}_{\VB}= t'_2 \, \tau\; , \qquad 
\la^{(1)}_{L}=\al \,\tau  \,,
\end{eqnarray}
where $t'_2$ is defined as
\begin{eqnarray} \label{eq:T2}
t'_2\,:=  \left. \frac{\de}{\de \tau }  \kappa  \int \de J \,p(J) \, \tanh^2\left( \frac{\be_c J}{1-\tau}\right) \right \vert_{\tau=0} >0 \,.
\end{eqnarray}

In all three cases the eigenvalue hence returns to positive values.
The instability of the paramagnetic saddle-point due to unstable
directions from the zeroth and first block are therefore cured by the
replica symmetric low-temperature solution. The ``dangerous'' direction
related to the replicon eigenvalue is contained in the $\R=2$ sector. 
Its detailed discussion requires a specification of the Hamiltonian 
which we therefore perform separately for the three case of interest.

\subsubsection*{The SK Model}
Using  $G_{\SK}(J)=\delta''(J)$ in (\ref{EV2}) for $\rho=2$ we find:
\begin{eqnarray}\label{eq:EV_rho_2_SK}
\la^{(2)} \,
\int \de \Th \, \Ph(\Th)\,\phi^{(2)} \big(\Th+h \big)
&=&\phi^{(2)}( h ) 
- 2 \beta^2\,\phi^{(2)}  \big(0 \big) \int \de \Th \, \Ph (\Th) \frac{1}{\cosh^4\bl \be(h +\Th )\br}\,.
\end{eqnarray}
At zero external field close to the transition temperature this equation 
can also be studied perturbatively  in the reduced temperature $\tau$. 
An expansion of the last equation at $h=0$ up to order $h^4$ turns 
the integral equation to a three dimensional eigenvalue problem.
One eigenvalue becomes negative:
\begin{eqnarray}\label{eq:ev_rho_2_SK}
\la^{(2)}=-\frac{4}{3}\;\tau^2 +\mathcal{O}(\tau^3),
\end{eqnarray} 
indicating the well-known instability of the RS solution. 

In the presence of an external field we identify  $\la^{(2)}=0$
with the instability-line, which starts at $T_c$ for $\exh=0$. 
The replica symmetric distribution of local fields is explicitly known 
for the SK model: 
\begin{equation}
  \Ph (h)=\frac{1}{\sqrt{4\pi q}}\exp\Le(-\frac{(h-\exh)^2}{4q} \Ri)\, ,
\end{equation}
where in the spin-glass phase  $q$ is the non-zero solution of 
\mbox{$q= \int \de h\, \Ph (h)   \tanh^2(\be h) $}.
Setting $\la^{(2)}=0$ and $h=0$ in (\ref{eq:EV_rho_2_SK}) we arrive at 
\begin{eqnarray}\label{eq:AT_SK}
1= 2\be^2
\int \frac{\de x}{\sqrt{2\pi}} \, 
\frac{\exp\big(-\frac{x^2}{2}\big )}{\cosh^4 \bl \be(\sqrt{ 2 q} x  +\exh) \br}  
\end{eqnarray}
which reproduces the famous AT-line for the SK-model in the \mbox{$\exh$-$T$}-plane 
 \cite{AT}.  

\subsubsection*{The VB Model}
In the case of the VB model the eigenvalue equation for $\R=2$ reads:
\begin{eqnarray}\label{eq:EV_rho_2_VB}
\la^{(2)} \,
\int \de \Th \, \Ph(\Th)\,\phi^{(2)} \big(\Th+h \big)
&=&\phi^{(2)}( h ) 
-\kappa \int \de \Th \, \Ph (\Th) 
\int \de J \, p(J)
\Le (\pa_h \pphi \Ri )^2( \Th+h,  J)\,
\phi^{(2)}\big( \pphi( \Th+h,  J) \big).
\end{eqnarray}
At zero external field an expansion of the eigenvalue equation (\ref{eq:EV_rho_2_VB})
can be invoked leading to the eigenvalue
\begin{eqnarray} \label{eq:ev_rho_2_VB}
\la^{(2)}&=&-\frac{1}{3}\, \frac{1+2\, t_4}{1-t_4}\,{ t'_2}^2  \;\tau^2 +\mathcal{O}(\tau^3)
\end{eqnarray}
with 
\begin{eqnarray} \nn
 t_4:= \kappa\,\int \de J \, p(J) \,\tanh^4(\be_c J)< \kappa \int \de J \,p(J)\, \tanh^2(\be_c J)=1\, , 
\end{eqnarray}
and $t'_2$ defined in eq.~(\ref{eq:T2}). Due to the inequality in
the last line $\la^{(2)}$ is negative below $T_c$ indicating the 
instability of the replica symmetric solution for this model.

Within the cavity approach the AT-line is described by  \cite{JEM}
\begin{eqnarray}\label{eq:EV_cavity}
 \mu \,\phi( h ) 
 =\kappa \int \de \Th \, \Ph (\Th) 
 \int \de J \, p(J)
\Le (\pa_h \pphi \Ri )^2( \Th+h,  J)\,
\phi\big( \pphi( \Th+h,  J) \big)\,,
\end{eqnarray}
where the RS phase becomes unstable when the largest
eigenvalue $\mu$ exceeds the value 1 \cite{rem}. Since
(\ref{eq:EV_rho_2_VB}) for $\la^{(2)}=0$ and 
(\ref{eq:EV_cavity}) for $\mu=1$ coincide we have reproduced the
stability criterion from the cavity method within the replica
approach also for non-zero external field.

The largest eigenvalue of eq.~(\ref{eq:EV_cavity}) can be
determined numerically by simple iteration. Unfortunately we do not
know about a similar straightforward method to determine the smallest
eigenvalue $\lambda^{(2)}$ of eq.~(\ref{eq:EV_rho_2_VB}).

\subsubsection*{The L\'evy spin glass}
The eigenvalue equation for the L\'evy SG is similar to the one for
the VB-model: 
\begin{eqnarray}\label{eq:EV_rho_2_Levy}
\la^{(2)} \,
\int \de \Th \, \Ph(\Th)\,\phi^{(2)} \big(\Th+h \big)
&=&\phi^{(2)}( h ) 
-C(\al)\int \de \Th \, \Ph (\Th) 
\int \frac{ \de J }{|J|^{\al+1}}
\Le (\pa_h \pphi \Ri )^2( \Th+h,  J)\,
\phi^{(2)}\big( \pphi( \Th+h,  J) \big)\, ,
\end{eqnarray}
where $C(\al)$ was defined in (\ref{defC}).
The expansion of the eigenvalue equation (\ref{eq:EV_rho_2_Levy})
to the leading order in the reduced temperature amounts to 
\begin{eqnarray}\label{eq:ev_Levy}
\lambda^{(2)}&=&
- \frac{\al^2}{3} \frac{1+2\,t_{4,\al}}{1-t_{4,\al}} \tau^2  + \mathcal{O}(\tau^3)\,,
\end{eqnarray}
with
\begin{eqnarray}\nn
t_{4,\al}:= C(\al)\int \frac{ \de J }{ |J| ^{\al+1}}\tanh^4( \beta_c J) < C(\al)\int \frac{ \de J }{ |J| ^{\al+1}}\tanh^2( \beta_c J)=1\, .
\end{eqnarray}
This proves that $\lambda^{(2)}$ is
indeed negative below $T_c$ and the replica symmetric phase is
unstable below $T_c$. 

The stability analysis for the L\'evy SG performed in  \cite{JEM}
using the cavity method gave rise to the equation 
\begin{eqnarray}\nn
 \mu \,\phi( h ) 
 =C(\al )\, \int \de \Th \, \Ph (\Th) 
 \int  \frac{ \de J }{|J|^{\al+1}}
\Le (\pa_h \pphi \Ri )^2( \Th+h,  J)\,
\phi\big( \pphi( \Th+h,  J) \big) \,,
\end{eqnarray}
where the instability of the RS solution was again signaled by
$\mu>1$ \cite{rem2}. In the presence of an external field we therefore
find  equivalence between the results obtained using the cavity and
the replica method.  

In  the  large connectivity limit of the VB model 
and in the SK-limit  ($\al \to 2$) of the L\'evy SG all our results
are consistent  with each other. 
In view of (\ref
{Gmodels}) we have  
$\lim_{\al \to 2 } G_{L}(J)= \delta''(J)= G_{\SK}(J)$. 
To obtain the large connectivity limit of the AT-line for
the VB model we use (\ref{eq:large_conn_SK}) 
and (\ref{eq:large_conn_Levy}) respectively in (\ref{eq:EV_rho_2_VB}). 
In the limit $N \to \infty$ the eigenvalue equation then acquires 
the desired form up to a constant depending on the  
details of the distribution $p(J)$ which can be absorbed in the  energy scale.

To see the equivalence  for the eigenvalues close to the
transition temperature  we use 
\begin{eqnarray}\label{eq:tc_limit}
t_2' \to \al \to 2  \quad {\rm and}
 \quad t_4 \to t_{4,\al}
 \to t_{4,2}  = \int \de J \,  \delta{''}(J) \, \tanh^4(\be_c J) = 0 \,
\end{eqnarray}
where the first arrow corresponds to the L\'evy limit of the VB model,
and the second one to the SK-limit of the L\'evy SG. 
If the second moment of the distribution $p(J)$ exists,
one obtains using (\ref{eq:large_conn_SK}) directly  $t_2' \to 2$  and $t_4 \to  0$.

\section{summary}\label{conc}
In the present paper we derived within the replica formalism
expressions for the AT-line of general mean-field spin-glasses
including strongly diluted and L\'evy spin glasses. Due to the
non-Gaussian character of the local field distribution in these models
an infinite number of order parameters is needed already at the
replica symmetric level. Following the approach of Monasson the 
fluctuations around the replica symmetric saddle-point are described
by an $2^n\times 2^n$ Hessian matrix. 

At high temperatures and in zero external field the distribution of local
fields is a delta-function and the determination of the 
eigenvalues of this Hessian is relatively straightforward. We find that 
all eigenvalues are positive at sufficiently high temperature and that
some of them tend to zero at the critical temperature, $T_c$, which
signals the transition to the glass phase. 

Below the critical temperature the RS order parameter develops a
non-trivial structure and the determination of the spectrum of the
Hessian becomes rather involved. However, using the symmetry of the
saddle-point under permutations of the replicas the Hessian can be
block-diagonalized and the sub-blocks relevant for the stability of RS
can be identified. Performing the $n\to 0 $ limit in these blocks 
turns the finite dimensional eigenvalue equations into integral
equations from which general expressions for the AT-lines may be
derived. 

We show the validity of our approach by applying it to three 
representative model systems: the SK-model as the standard model with 
Gaussian field distribution, the VB-model as the standard model for
diluted spin-glasses for which higher moments of the field
distribution are essential, and the L\'evy spin glass as standard
model for spin-glasses with diverging second moment of the coupling
distribution. 

We believe that with our stability analysis the replica-symmetric
theory of general spin-glass models is now complete. 

\begin{acknowledgments}
We would like to thank  Marc M\'ezard, Remi Monasson and  Martin Weigt for interesting
discussions. Financial support from the Deutsche Forschungsgemeinschaft
under EN 278/7 is gratefully acknowledged.
\end{acknowledgments}

\appendix
\section{The spectrum of $\He^{\RR}$}\label{}
First of all we show that the eigenvalues of $\He^\RR_{\rm
  Sym}$ and $\He^\RR$ coincide. To this end we note that the
  symmetrization operators $\mathcal{S}_{n-2\R}(\s)$ and
  $\mathcal{A}_{2\R} $ can be represented as appropriate combinations
  of $D(\pi), \,\,\pi \in \Sigma_n$. The operator
  $\mathcal{S}_{n-2\R}(\s)$, as a sum of ${n-2\R} \choose {\s-\R} $
  permutations, and  $\mathcal{A}_{2\R}$ as a product  
\begin{eqnarray}
\mathcal{A}_{2\R} = \prod_{a=1}^\R \Le [D(e)-D\big((2a,2a-1)\big) \Ri],
\end{eqnarray} 
where $e$ denotes the identity of the group and $(2a,2a-1)$ the
transposition of the elements $2a$ and $2a-1$. Being elements of the
group algebra all operators $\mathcal{S} $ and $\mathcal{A}$
commute with $\He$. Therefore  
\begin{eqnarray}
\lr \s;\R \Ri | \He \Le|   \T  ; \R  \rl
= 
\LS \s \SL
\He \, \mathcal{A}_{2\R}\,
\mathcal{S}_{n-2\R}(\s)       \Le|   \T  ;\R \rl \,. 
\end{eqnarray}
The vector $ \Le|   \T;\R  \rl$ is symmetric in the last $n-2\R$
entries. The action of any $D(\pi)$ acting only on these last entries
is hence trivial and we find   
\begin{eqnarray} 
 \mathcal{S}_{n-2\R}(\s)   \Le|   \T ;\R \rl  = {{n-2\R} \choose {\s-\R} } \Le|   \T ;\R \rl,
\end{eqnarray}
since $\mathcal{S}_{n-2\R}(\s)$ is a sum of  ${n-2\R} \choose {\s-\R}
$ permutations. From $\Le [  D(e)-D\big((1,2)\big) \Ri ] \,
\big(\p \m-\m\p \big) = 2  \big(\p \m-\m\p \big)\,,$ 
one similarly derives $ \mathcal{A}_{2\R}   \Le|   \T ;\R \rl  = 2^\R \Le|   \T ; \R \rl$.
The action of $ \mathcal{A}_{2\R} \mathcal{S}_{n-2\R}(\s)$ reproduces $ \Le|   \T;\R  \rl$
up to the constant $ 2^\R {{n-2\R} \choose {\s-\R}}$ which is the squared norm of $\Le| \s ;\R \rl $.
One hence has the following relation between the matrix elements of $\He$ and $\He^{\RR}$:
$ \lr\s; \R \Ri |\He \Le|   \T ;\R  \rl = \lr \s ;\R \Ri.  \Le| \s ;\R   \rl \,\He^\RR ( \s ,\T)$.
Using this relation in the  eigenvalue equation for $\He^{\RR}_{ \rm Sym} $:
\begin{eqnarray} \nn
\la^{\RR} \tilde \chi^\RR(\T)
&= &\sum_{\s=\R}^{n-\R}  \He^\RR_{\rm Sym}(\s ,\T ) \,  \tilde \chi^\RR(\s) 
 =  \sum_{\s=\R}^{n-\R}  \sqrt{ \frac{ \lr \s;\R \Ri. \Le| \s ; \R \rl}{   \lr \T;\R \Ri. \Le|   \T;\R  \rl   }}\,\,
\He^\RR ( \s ,\T)  \,   \tilde \chi^\RR(\s)\\ \nn
\Leftrightarrow \qquad  \la^{\RR} \chi^\RR(\T)
&=& \sum_{\s=\R}^{n-\R}  \He^\RR ( \s ,\T)  \,    \chi^\RR(\T),
\qquad
 \rm{ with  }
\qquad
\chi^\RR(\s)=  \sqrt{   \lr \s;\R \Ri. \Le| \s; \R  \rl} \,\tilde \chi^\RR(\s)
\end{eqnarray}
we see that the eigenvalues of the matrices coincide. 
\section{The limit $n \to 0$ }\label{App:n_to_0}
To perform  the limit  $n \to 0 $ we switch to the characteristic
functions defined in (\ref{n20}). Transforming the whole eigenvalue
equation (\ref{EV1}) one has to calculate the quantity
\begin{eqnarray}\label{trHe}
\sum_{\T=\R}^{n-\R} \He^\RR(\s ,\T ) \exp \Le( \be h  (2 \T -n ) \Ri)
\end{eqnarray}
which can be performed without the explicit evaluation of the matrix
elements $\He^\RR(\s ,\T )$. As we will see  the transformed
quantities will allow for a continuation to real $n$.  
The $n\to 0 $ limit then turns the finite dimensional eigenvalue equation into
the  integral equation (\ref{EV2}).

In section \ref{GT} the entropic and the energetic parts of the Hessian,
$\He_S$ and $\He_E$, were decomposed into a tensor product of identical
$2\times 2$ matrices. Therefore we first calculate the transformation
(\ref{trHe}) for a tensor product of n $2\times 2$ matrices $\G$ and
use the general result for the two parts of the Hessian. 
\begin{eqnarray} \label{g-transf}
&&\sum_{\T = \R}^{n- \R}
\LS \s \SL
    \TP_{a=1}^n  \G  \Le |  \T ; \R \rl
\exp \big( \be h (2 \T -n) \big)
\\ \nn
 && =
\sum_{\tau = \R}^{n- \R}\big( \bp \bm \,\big )^{\R}
\big( \bp     \, \big)^{\s-\R}
\big( \bm     \, \big)^{ n-\s-\R}
\TP_{a=1}^n  \G
\big( \p \m - \m \p \big)  ^\R \Le| \T-\R \rl_{n-2\R} 
\exp \big( \be h(2\T -n ) \big)\\ \nn
&&=\Big( \bp\bm \G \otimes \G   \big( \p \m - \m \p  \big )   \Big )^\R
\sum_{\tau = \R}^{n- \R}  \big( \bp \big)^{\s-\R} \big( \bm \big)^{n-\s-\R}
\TP_{a=1}^{n-2\R}  \G
\Le [ \sum_{\sum \T_a =  2 \T-n} \TP_{a=1}^{n-2 \R} \Le|\T_a \rl \Ri]
\exp \Le( \be h \sum_{a=\R}^{n-\R}\T_a  \Ri)
\\ \nn
&&=
\Big( \bp \G \p \bm \G \m - \bp \G \m  \bm \G \p \Big)^\R
\sum_{ \{\T_a = \pm 1 \}}
\Le [
\prod_{b=1}^{\s-\R}    \bp \G \Le|\T_b\rl e^{\be h\T_b}
\prod_{c=1}^{n-\s-\R}  \bm \G  \Le|\T_c\rl e^{\be h\T_c} 
\Ri ]
\\ \nn
&&=
\Big( \G_{++}  \G_{--}  - \G_{+-} \G_{-+} \Big)^\R
\Le(\sum_{ \T_b = \pm }  \bp \G \Le|\T_b\rl e^{ \be h\T_b}\Ri)^{\s-\R}
\Le( \sum_{ \T_c = \pm } \bm \G  \Le|\T_c\rl e^{\be h\T_c} \Ri)^{n-\s-\R }
\\ \nn
&&=
\Big( \G_{++}  \G_{--}  - \G_{+-} \G_{-+} \Big)^\R
\Big( \G_{++} e^{\be h}+ \G_{+-} e^{-\be h}\Big)^{\s-\R}
\Big( \G_{+-} e^{\be h}+ \G_{--} e^{-\be h}\Big)^{n-\s-\R},
\end{eqnarray}    
where $\G_{\s \T}:=  \lr \s \Ri| \G \Le | \T \rl, \, \s,\T \in \pm$  denote 
the matrix elements of $\G$.
In the second line we have used the definitions of $\LS \s \SL$
and $  \Le |  \T ; \R \rl$ and then decomposed the
symmetric part of  $  \Le |  \T ; \R \rl$  as in (\ref{sympart}),
using $2\T-n = \sum_{a} \T_a$. 

The entropic contribution (\ref{tensor_decomp_entrop}) can be written
as 
\begin{eqnarray}
\He_S 
= \int \frac{\de r \de r' }{2 \pi} \frac{e^{irr'}}{  c(r)}
  \TP_{a=1}^n  \exp \Le(-ir' \Pa_3 \Ri)  
= \int \frac{\de r \de r' }{2 \pi} \frac{e^{irr'}}{  c(r)}
  \TP_{a=1}^n  \G^S   . 
\end{eqnarray}
The entropic matrix $\G^S$ is diagonal:  $ \G^S_{++} =e^{-ir'}, \quad
\G^S_{--}=e^{+ir'} $, and using (\ref{g-transf}) we find 
\begin{eqnarray}\label{ENTRO} \nn
\sum_{\T=\R}^{n-\R} \He^\RR_S(\s ,\T ) \exp \bl \be \,h \, (2 \T -n ) \br
&=&
\int \frac{\de r \,\de r' }{2 \pi}\, \frac{e^{irr'}}{  c(r)}
\;\sum_{\T=\R}^{n-\R}\;
\LS \s \SL \, \TP_{a=1}^n \, \G^S(r') \,       
  \Le |  \T; \R \rl \, \exp \bl \be h  (2 \T -n ) \br \\ \nn
&&\hs{-5cm}=\int \frac{\de r \de r' }{2 \pi} \frac{e^{irr'}}{ c(r)}
\big(\G^S_{++}e^{\be h} \big)^{\s-\R} \big( \G^S_{--}e^{-\be h} \big)^{n-\s -\R}
=\int \de r \frac{1}{c(r)}\D\bl r-(2\s-n)\br \,
\exp\big(\be h(2 \s -n)\big) \\ \nn
&&\hs{-5cm}=
\sum_{\T=\R}^{n-\R}\frac{ \D_{\s,\T} }{ c( 2\s-n   )}\exp\big(\be h(2
\T -n)\big) \; .
\end{eqnarray}
The expression in the last line leads to eq. (\ref{HeRhoS}). Due to
the structure  of the RS solution (\ref{RS}) it is convenient to
determine the inverse of $\He^\RR_S$  which amounts to :
\begin{eqnarray}\label{He_S}
{ \Le( \He^\RR_S \Ri)}^{-1}( \s ,\T)  
=  \D_{\s,\T}\, c( 2\s-n   )
= \D_{\s,\T}\,
\int  \de \Th \, \Ph ( \Th)  
\frac{  \exp\big(\be \Th(2\T-n)\big)  }{\big(2   \cosh(\be \Th)\big)^n } \,.
\end{eqnarray}
The energetic contribution (\ref{tensor_decomp_energ}) reads
\begin{eqnarray}
\He_E
=-\int \de J \, G_M(J) \TP_{a=1}^n  \Le [ e^{\be  J}\Pa_0+e^{-\be  J}
\Pa_1 \Ri ] =-\int \de J \, G_M(J) \TP_{a=1}^n  \G^E (J),
\end{eqnarray}
with diagonal elements  $\G^E_{\s\s}(J)= e^{\be J}$, and off-diagonal terms
$\G^E_{\s-\s}= e^{-\be J}, \, \,  \s= \pm 1$. Using (\ref{g-transf})
we get 
\begin{eqnarray}\label{ENE}
-\sum_{\T=\R}^{n-\R} \He^\RR_E(\s ,\T ) \exp \big( \be h (2 \T -n ) \big)
&=&
 \int\de J  \,G_M (J) \, \sum_{\T=\R}^{n-\R}
\LS \s \SL 
       \TP_{a=1}^n  \G^E (J)    
 \Le |  \T; \R \rl \exp \big( \be h (2 \T -n ) \big)    \\\nn
&&\hs{-4cm}
=
 \int\de J  \,G_M (J) \,  
 \Big(  e^{ 2 \be  J}      - e^{- 2 \be  J}     \Big)^{\R}
 \Le(  e^{  \be ( J+ h)} + e^{ -\be ( J +h )} \Ri)^{\s-\R}
 \Le(  e^{  \be (- J+ h)} + e^{ -\be (-J +h )} \Ri)^{n-\R-\s}\\ \nn
&&\hs{-4cm}
=
\int \de J  \,G_M (J) \, \big( \pa_h \pphi(h,  J) \big)^\R  \,  w_n(h,J)\,
 \exp \big (\be\,(2\s-n)\, \pphi(h,  J)   \big) \\ \nn
\end{eqnarray}
with
\begin{equation*}
w_n(h,J) = \big( 2 \cosh\be ( J +h ) 2 \cosh\be ( J -h )
       \big)^{\frac{n}{2}}
\end{equation*}
and 
\begin{equation*}
\pa_h \pphi(h,  J)=
= \frac{1}{\cosh^2(\be h )}  \frac{\tanh(\be J)}{ 1-\tanh^2(\be J)\tanh^2(\be h)}\,.
\end{equation*}
\section{The eigenvalue equation}\label{App:ev_eq}
We start from the eigenvalue equation (\ref{EV1}) for the  matrices
$\He^\RR$. Each of the eigenvalues corresponds to one of the $n+1-2\R$
representations $D^\RR$ which arise in the decomposition
(\ref{D-decomp}). Splitting the eigenvalue equation into energetic
and entropic part leads to  
\begin{eqnarray}\label{EV3}
\la^\RR   \chi^\RR(\T)
 =   \sum_{\s=\R}^{n-\R}\He^\RR_S(\s, \T)  \chi^\RR(\T) 
    +\sum_{\s=\R}^{n-\R}\He^\RR_E (\s, \T) \chi^\RR (\s)
\end{eqnarray}
which is equivalent to 
\begin{eqnarray}\nn
\la^\RR  \int   \frac{  \de \Th \, \Ph ( \Th)  }{(2   \cosh(\be \Th))^n }  \chi^\RR(\T) \exp(\be \Th(2\T-n)) 
 &=& \chi^\RR(\T)
+   \int   \frac{  \de \Th\,  \Ph ( \Th)  }{(2   \cosh(\be \Th))^n }
 \sum_{\s=\R}^{n-\R}\
 \He^\RR_E (\s, \T)   \exp(\be \Th( 2\T-n))
 \chi^\RR (\s)\;. 
\end{eqnarray}
Here we have used the explicit form of the inverse of $\He^\RR_S $
given in (\ref{He_S}). We now perform the transformations explained
above which allows to perform the limit  $n\to 0 $. The transformation
of the l.h.s. of the last equation reads 
\begin{eqnarray} \nn
\la^\RR  \int  \de \Th \, \Ph ( \Th) \,  \frac{1}{ \big(2 \cosh (\be \Th) \big )^n  }    \sum_{\T=\R}^{n-\R}  
 \exp \big( \be (\Th+h) (2 \T -n) \big) \chi^\RR(\T)
=\la^\RR  \int \de\Th  \frac{1 }{ \bl 2 \cosh (\be \Th  )\br ^n}\Ph( \Th)\, \phi^\RR(\Th+h)
\end{eqnarray}
whereas the r.h.s amounts to :
\begin{eqnarray} \nn
&& \sum_{\T=\R}^{n-\R}   \chi^\RR(\T)\exp \big(\be h( 2\T-n)\big) 
+ \int  \de \Th \, \Ph ( \Th)\frac{1}{ \big(2 \cosh( \be  \Th )\big)^n  } 
  \sum_{\T,\, \s=\R}^{n-\R}\
 \He^\RR_E (\s, \T)
\exp\big(\be (\Th+h)( 2\T-n)\big) \chi^\RR (\s) 
\\ \nn &&
=\phi^\RR(h)
+ \int  \de \Th \, \Ph ( \Th) \frac{ 1 }{ \big(2 \cosh(\be \Th  )\big)^n}     \, 
 \sum_{ \s=\R}^{n-\R}\
 \Le( \sum_{ \T=\R}^{n-\R} \He^\RR_E (\s, \T)
\exp\big(\be (\Th+h)( 2\T-n) \big) \Ri)
 \chi^\RR (\s) 
\\ \nn &&
=\phi^\RR(h)
-\int\de \Th \,  \Ph ( \Th) \frac{1  }{ \bl 2 \cosh(\be \Th  ) \br^n}
  \int \de J  \,G_M (J) \, \Le( \pa_h \pphi(\Th+h,  J) \Ri)^\R w_n(\Th+h,J)\,\phi^\RR\big(\pphi(\Th+h,J)\big ),
\end{eqnarray}
where we have used (\ref{ENE}). The limit $n\to 0$ may now be
performed which yields our central result (\ref{EV2}). 



\begin{thebibliography}{99}
\bibitem{JEM1}A similar expansion of an eigenvalue problem was recently used in  \cite{JEM}.
\bibitem{rem} To obtain this expression the substitution $g_i \mapsto
  g_i-h_i$ has to be performed in eq.~(39) 
of \cite{JEM}.
\end{thebibliography}
\end{document}